# Mode couplings and resonance instabilities in dust clusters

Ke Qiao,[1] Jie Kong,[1] Eric Van Oeveren,[2] Lorin S. Matthews[1] and Truell W. Hyde[1]

[1]Center for Astrophysics, Space Physics and Engineering Research, Baylor University, Waco, Texas 76798-7310, USA

[2]The Physics Department, University of Wisconsin-Milwaukee, Milwaukee, Wisconsin 53201, USA

The normal modes for three to seven particle two-dimensional (2D) dust clusters in a complex plasma are investigated using an N-body simulation. The ion wakefield downstream of each particle is shown to induce coupling between horizontal and vertical modes. The rules of mode coupling are investigated by classifying the mode eigenvectors employing the Bessel and trigonometric functions indexed by order integers ($m$, $n$). It is shown that coupling only occurs between two modes with the same $m$ and that horizontal modes having a higher shear contribution exhibit weaker coupling. Three types of resonances are shown to occur when two coupled modes have the same frequency. Discrete instabilities caused by both the first and third type of resonances are verified and instabilities caused by the third type of resonance are found to induce melting. The melting procedure is observed to go through a two-step process with the solid-liquid transition closely obeying the Lindemann criterion.

## I. INTRODUCTION

Plasma crystals have been studied extensively since their experimental discovery in 1994 [1-3]. In ground-based experiments, plasma crystals are typically formed in the sheath region above the lower electrode of a rf discharge plasma. The dust particles composing these systems become negatively charged and are then levitated by the balance between gravity and the electrostatic force. Due to their characteristic length scale of hundreds of microns and time scale of tens of milliseconds, these systems have proven themselves useful as model systems for research on fundamental processes such as waves (or collective oscillations) in classical fluids and solids at the kinetic (atomistic) level.

In two-dimensional (2D) plasma crystals consisting of several thousand dust particles, in-plane [4-6] and out-of-plane [7, 8] dust lattice wave (DLW) modes have been studied. In-plane waves include longitudinal and transverse modes, both of which exhibit acoustical characteristics. Out-of-plane waves are transverse and have optical characteristics. In an ideal system in thermal equilibrium, these modes are independent, and the 2D crystalline structure will remain stable until the out-of-plane wave mode reaches a minimum frequency of zero, causing a one- to two-layer transition [9].

Plasma crystals are highly thermal non-equilibrium systems with continuous energy input from the ambient plasma. The most important mechanism of this sort is the ion flow within the sheath, directed towards the negatively charged lower electrode, which creates an ion focusing area with positive space charge below the dust particle, i. e. the ion wakefield [10-14]. It was recently predicted that the ion wakefield can induce couplings between in-plane longitudinal and out-of-plane transverse waves, causing resonance instabilities when the two mode frequencies approach each other [15-19]. This has since been proven experimentally [20-22].

This phenomenon raises the interesting question of whether the same physics might exist within a dust cluster consisting of only a few particles. Dust clusters are finite systems and therefore manifest unique characteristics, for example, exhibiting ordered structure such as concentric shells of particles [e.g. 23, 24]. As such, their dynamic properties are no longer described as wave modes, but are instead defined in terms of normal modes having discrete frequencies. While the normal modes in dust clusters have been examined in detail [23-31], to our knowledge the couplings between these modes have only been investigated for two- and/or three-particle systems [32-35], and the relationship between these couplings and those found in plasma crystals has not yet been examined.

Melting and/or the solid-liquid transition, a concept originally defined for larger physical systems, has recently become a topic of increasing interest for finite systems across various research fields [36, 37]. This solid- and liquid-like behavior has been observed across a wide variety of finite systems, for example electrons in quantum dots [38], ions in traps [39], and atomic clusters [40, 41]. A theoretical generalization of the definition of melting for small particles systems was proposed by Proykova and Berry in terms of the (thermodynamic) phase space of the particles [41]. However, there are still many open questions. Melting of dust clusters in complex plasmas has been investigated, but only when induced by laser heating [30, 31, 42] or excited by the existence of a single particle below the cluster [43]. Both of these are completely different mechanisms from the mode couplings created by the ion wakefield.

A detailed analysis of mode couplings and related processes such as instabilities and melting in dust clusters containing more than three particles is still outstanding. Since analyses of experimentally observed mode couplings and melting [20-22] are based on investigations of the power spectra, a numerical simulation for these systems is sorely needed. Such a simulation would allow dust particles to be tracked at the kinetic level and the power spectra obtained in the same manner as it is experimentally. This would in turn allow direct comparison with experimental results.

In this paper, previous numerical simulation research on ion wakefield-induced mode couplings for two-particle systems [35] is extended to dust particle clusters consisting of three to seven particles. Both mode coupling and subsequent resonance instabilities are investigated for various cluster sizes and external confinement conditions. Rules of coupling are deduced and instability-induced melting is analyzed.

## II. MODEL AND METHOD

In this work, the ion wakefield is modeled by adaptation of the point charge model. This model is commonly used [15-17, 32, 35] and has been effectively employed to explain the structure and dynamics of both dust crystals and particle pairs in plasma [44-46]. The point charge model assumes that the ion wakefield around a dust particle with negative charge $Q$ takes the form of a potential created by the particle itself and a positive point charge $q$ located at a distance $l$ beneath it (Fig. 1). It provides a reasonable analytical approximation while offering the advantage of highlighting qualitative features of the physical process, such as the positive space charge effect beneath the dust particle.

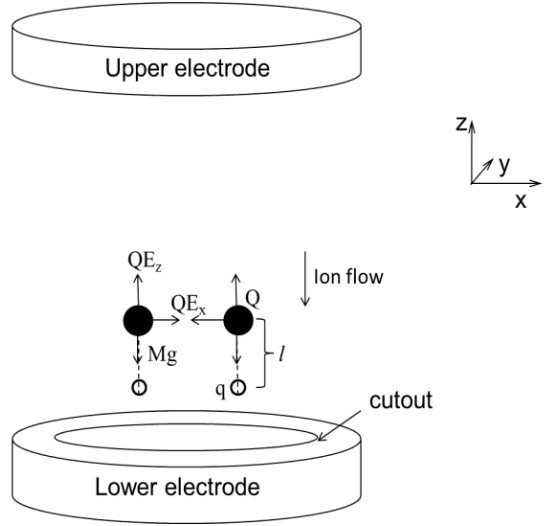

FIG. 1. The point charge model for the ion wakefield employed for a typical experimental setup, where $Q$ is the dust particle charge, $q$ is the point charge, $l$ is the distance between the dust particle and the point charge, $Mg$ is gravity, and $E_x$ and $E_z$ are the horizontal and vertical electric fields, respectively.

Employing the point charge model, the interaction between two dust particles suspended in a complex plasma is due to a combination of a screened Coulomb potential (the Debye–Hückel or Yukawa potential) and the point charge potential [32, 35], given by

$$U(r) = Q \frac{\exp(-|r - r_Q|/\lambda)}{|r - r_Q|} - q \frac{\exp(-|r - r_q|/\lambda)}{|r - r_q|} \qquad (1)$$

where $r_Q$ and $r_q$ are the real and virtual particle positions respectively, and $\lambda$ is the screening length. In the sheath region, where the majority of dust particles levitate in experiments on Earth, the ion-flow velocity is such that the mean kinetic energy of the ions equals or exceeds that of the electrons [14]. For this case, experimental measurements have shown the screening length to be roughly equal to the bulk-electron Debye length [47, 48]; thus, screening is often attributed to the electrons rather than the ions [10-12]. However, due to the complexity of the plasma sheath, the origin of this screening remains an open question, and will not be specified here. The resulting interaction between the particles now includes the dust-dust interaction, the interaction between dust particle $i$ and virtual particle $j$, and the interaction between dust particle $j$ and virtual particle $i$. This interaction is obviously non-Hamiltonian.

In a typical experiment on earth, dust particles are assumed to be confined in the vertical direction within the potential well formed by the electrostatic field and gravity, and in the horizontal direction by the electric field produced by the experimental setup [49]. In this case, the external forces acting on the particles are

$$\begin{aligned} F_{z,ext} &= E_z(x, y, z)Q - Mg \\ F_{x(y),ext} &= E_{x(y)}(x, y, z)Q \end{aligned} \qquad (2)$$

where $x$, $y$ and $z$ are representative particle coordinates. The electric field is assumed linear for small oscillations; thus $E_z = E_0 + E_z'z$, $E_x = E_x'x$ and $E_y = E_y'y$. where $E_0$ is the vertical electric field at the equilibrium position. This research is focused on determining the effects created by the ion wake field; other mechanisms, such as charge variation are ignored. (Such charge variations do not change the results qualitatively but does renormalize the mode couplings [17].) For this case, the particle charge $Q$ remains constant and the forces given in Eq. 2 constitute a parabolic potential well in both the vertical and horizontal directions [49, 50].

An N-body code, box_tree [8, 9, 28, 35, 51-54], was employed to model the dynamics of the dust system. Particles are trapped within a parabolic potential well in the horizontal direction and by a linear electric field and gravity in the vertical direction (see Eq. 2). Given the small number of particles modeled, the force on each particle exerted by all other particles is calculated directly from the potential given in Eq. 1 with no approximations. A simple Debye–Hückel potential (the first term on the RHS of Eq. 1) is used to calculate the forces between grains without the ion wakefield present.

All simulations begin from an initial random distribution of particles within a box of $10 \times 10 \times 10$ mm$^3$. The center of mass of the particle system is located at the center of the box, which is also defined as the origin of the coordinate system. Simulations were conducted for dust clusters consisting of three to seven particles. Clusters with $N > 7$ were not considered since at higher particle numbers, instabilities resulting from mode couplings overlap and become indistinguishable, and the existence of multiple metastable states adds additional complexity. Simulation parameters represent normative experimental values, with a particle diameter $d = 8.89$ μm, a particle mass $M = 5.56 \times 10^{-15}$ kg, particle charge $Q = 2.4 \times 10^{-15}$ C (15000e) and screening length $\lambda = 300$ μm. The ion wakefield is modeled by assuming a point charge $q$ of $Q/4$ or $Q/8$, located a distance $l = \lambda/2$ below the dust particles. The radial confinement in the horizontal direction ($E_x' = E_y'$) is fixed at $3.655 \times 10^4$ V/m$^{-2}$ ($\omega_x = \omega_y = 12.5$ Hz). The vertical confinement $E_z'$ is allowed to vary from ~$5 \times 10^4$ to $2.3 \times 10^5$ V/m$^{-2}$ ($\omega_z \approx 14.7$ ~ $31.5$ Hz).

Under these conditions, the particle system is first cooled through friction as defined by $F_f = -\mu M \mathbf{v}$, where $\mathbf{v}$ is particle velocity and $\mu = 10$s$^{-1}$. The system is also established in a heat bath created by elastic random collisions between the dust particles and the ambient particles [28]. The ambient temperature is set to be 1 K, allowing mutual particle interactions to dominate the inherent thermal motion. Friction is removed after formation of a stable cluster configuration (elapsed simulation time of 15s) and the system remains in the heat bath. Energy checks were run for dust systems without an ion wakefield. Figure 2 shows the temporal evolution of the average kinetic energy for a five particle system. It can be seen that the system initially exhibits very low kinetic energy, as would be expected due to frictional cooling, and then gradually increases until saturation at around $1.8 \times 10^{-23}$ J, which corresponds to a temperature of 0.87 K. At this point, the cluster not only exhibits a stable configuration but remains thermally stable within a fluctuation in kinetic energy of $\pm 0.6 \times 10^{-23}$ J. No instability was observed across all particle numbers tested and over the entire parameter range. Thus, the instabilities observed by this research for dust clusters are induced via the ion wakefield.

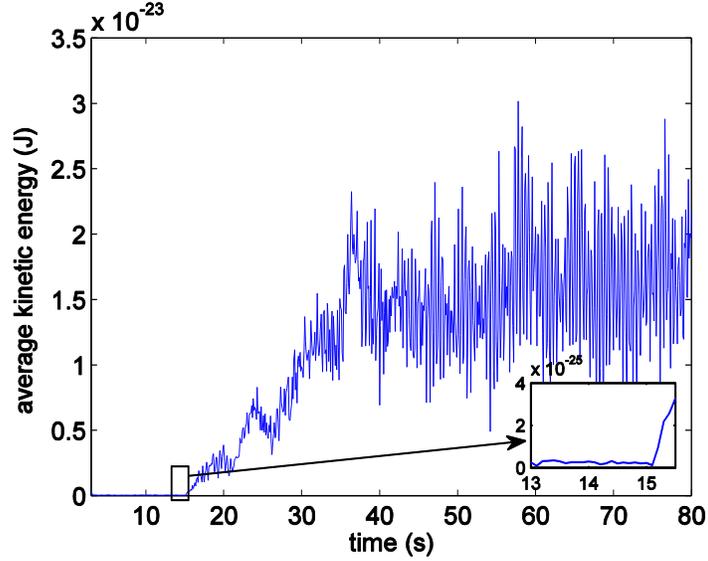

Fig. 2. (Color online) Temporal evolution of the average kinetic energy for a five particle system without the ion wakefield with $E_z'=1.705\times10^5$V/m$^{-2}$.

For thermally stable clusters, particle oscillations were tracked for approximately 67 s with each particle's position and velocity saved at intervals of 1/15 s. Mode spectra were obtained employing the technique described by Melzer [25], where the time series of the particle velocities is projected onto the direction of the eigenvectors for each pure mode (i.e. modes obtained while disregarding the ion wakefield). The mode power spectrum was then obtained through Fourier transformation.

## III. RESULTS

Couplings between the normal modes of a three-particle cluster were examined initially. In agreement with previous research [25, 28], six types of pure modes were identified for a three particle cluster: four types of horizontal (in-plane) modes, the rotation mode, sloshing modes, "kink" modes and the breathing mode, and two types of vertical (out-of-plane) modes, the vertical sloshing mode and vertical relative modes. Due to symmetry in the *x*- and *y*-directions, the horizontal sloshing, "kink", and vertical relative modes all have a degeneracy of two. The eigenvectors of these modes are shown in Fig. 3.

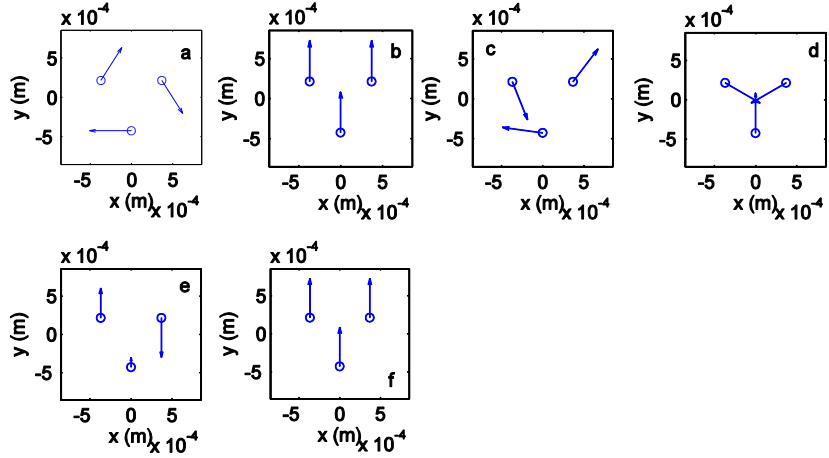

Fig. 3. (Color online) Eigenvectors for the normal modes in a three particle cluster: (a) the horizontal rotation mode (mode 1), (b) the horizontal sloshing mode (modes 2 and 3), (c) the horizontal kink mode (modes 4 and 5), (d) the horizontal breathing mode (mode 6), (e) the vertical relative mode (modes 7 and 8) and (f) the vertical sloshing mode (mode 9) .

The power spectra for a three particle cluster are shown in Fig. 4 with the vertical confinement set at $E_z' = 1\times 10^5$ V/m$^{-2}$, where modes numbered 1-6 correspond to the horizontal modes while modes 7 to 9 correspond to the vertical modes, as shown in Fig. 3. Without the ion wakefield, the power spectrum shows that only one mode (two if there is a degeneracy) is excited at a given spectral frequency (Fig. 4(a)); thus all modes are pure modes, i. e. there is no coupling between them. With the introduction of the ion wakefield, more than one mode (more than two if there is a degeneracy) can appear for some spectral frequencies (Fig. 4(b)). For example at the spectral frequency of the breathing mode (mode 6), $f = 4.6$ Hz, not only the breathing mode itself but also a light spectral line at mode number 9 (the vertical sloshing mode) can be seen. This means that this mode is no longer a pure horizontal breathing mode, but is modified and acquires mixed polarization, i.e., the horizontal breathing mode is now coupled to the vertical sloshing mode. Following similar arguments one can see that there are two additional couplings between the horizontal and vertical modes: the coupling between the horizontal "kink" modes (modes 4 and 5) and the vertical relative modes (modes 7 and 8) and between the vertical relative and horizontal sloshing modes (modes 2 and 3). It can also be seen that the resulting mode frequencies are slightly lower compared to those without the ion wakefield (see inset in Fig. 4(b)), in agreement with [35], where a two particle cluster was investigated.

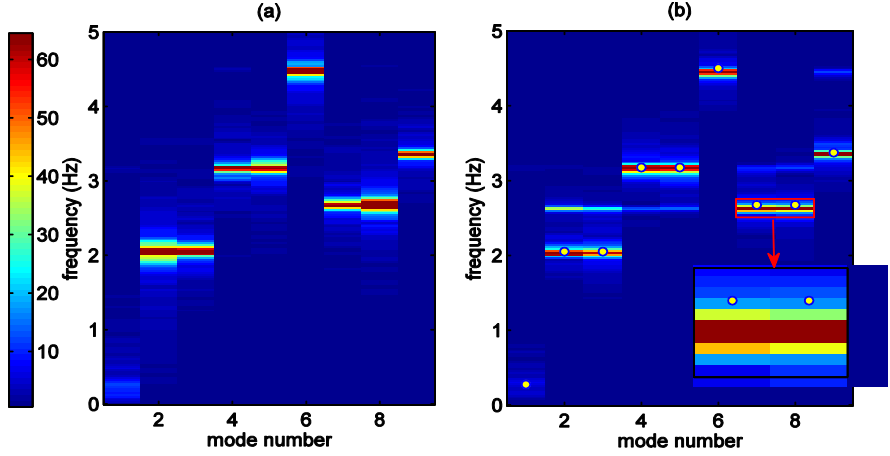

Fig. 4. (Color online) Power spectra for a three particle cluster with $E_z' = 1 \times 10^5$ V/m$^{-2}$ (a) without and (b) with ion wakefield. The dots in (b) correspond to spectral frequencies without the ion wakefield. The color (grayscale) bar represents the power spectral density in units of $(m/s)^2/Hz$.

As the vertical confinement $E_z'$ is varied, vertical mode spectral lines shift in frequency. In this case, resonances occur whenever two coupled modes (one horizontal and one vertical) have equal frequencies. Plotting the mode frequencies as functions of $E_z'$ (the $E_z'$-fr functions), these resonance points correspond to the points where two function curves cross (Figures 5 – 9). (Note that uncoupled modes do not affect each other; thus, the intersection points of their $E_z'$-fr functions do not correspond to resonances.) Resonances can be classified into one of three types, as illustrated in figures 5, 6 and 7.

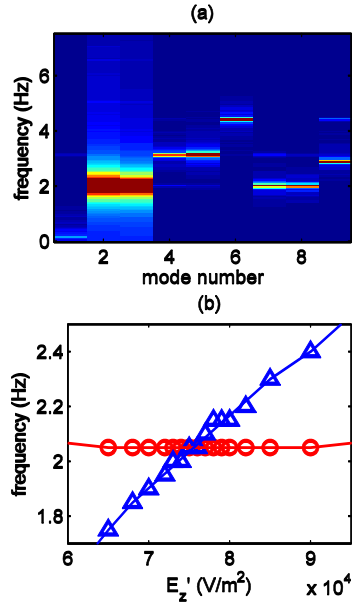

Fig. 5. (Color online) The first type of resonance illustrated by (a) the power spectra and (b) $E_z'$-fr functions around the resonance points. The results shown are for a three particle cluster at the resonance between the vertical relative (blue triangles) and horizontal sloshing (red circles) modes. The decoupled shapes of the $E_z'$-fr functions are represented by the solid lines.

The first type of resonance occurs if and only if one of the two coupled modes is a horizontal or vertical sloshing mode and exhibits a unidirectional characteristic. That is, the mode coupled to the sloshing mode has mixed polarization with a sloshing component, while the sloshing mode remains a pure mode. This type of resonance causes an instability, which is seen by an increase in both spectral energy density and particle velocities. However, due to its unidirectional characteristic, this instability occurs only for the sloshing mode. As an example, Fig. 5(a) shows the power spectrum for a three particle cluster with $E_z' = 8 \times 10^4$ V/m$^{-2}$, corresponding to the resonance between the vertical relative and horizontal sloshing modes. Instabilities can be clearly identified by an increased energy density at modes 2~3 (the horizontal sloshing modes), which is much larger than all other modes. On the other hand, modes 7~8 (the vertical relative modes) do not exhibit any increase in energy density. Another characteristic of this type of resonance is that the $E_z'$-$fr$ functions in the vicinity of the resonance point show no deviation from their decoupled shapes (see solid lines in Figs. 5(b), 6(b), 7(b)). The $E_z'$-$fr$ functions for the coupled modes are shown by triangles/circles and are obtained from the numerical simulation. The shape of a decoupled mode is obtained by connecting the data points for this mode far away from the resonance point on both sides. In a two particle cluster, this is the only type of resonance that exists [35].

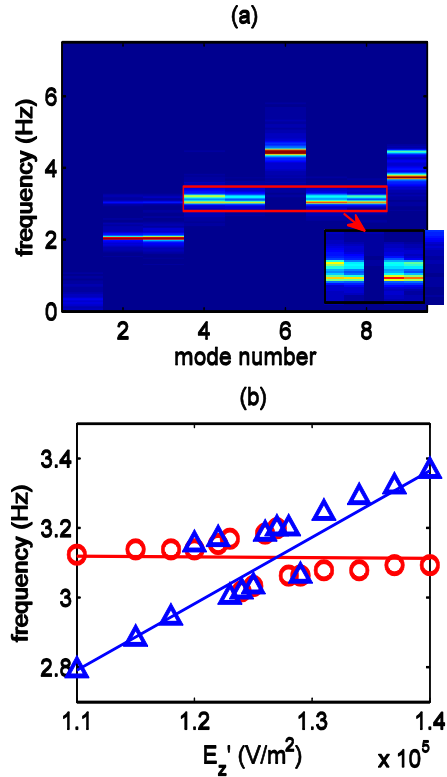

Fig. 6. (Color online) The second type of resonance illustrated by (a) the power spectra and (b) $E_z'$-$fr$ functions around the resonance point for a three particle cluster at the resonance between the vertical relative (modes 7 and 8, blue triangles) and horizontal "kink" (modes 4 and 5, red circles) modes. The decoupled shapes of the $E_z'$-$fr$ functions are represented by the solid lines.

The second type of resonance occurs between two modes having mixed polarization. In Fig. 6 the resonance between the vertical relative and horizontal "kink" modes in a three particle cluster is shown. In this case, no instability exists as shown by the lack of increase in energy density in either of the two modes. Instead, both modes show a double spectral line structure (see inset in Fig. 6), the $E_z'$-fr functions in the vicinity of the intersection point (Fig. 6(b)) deviate from their decoupled shapes, showing a reconnection between modes for lower and higher $E_z'$ and a forbidden frequency gap appears. All of these are characteristic of a linear mode conversion.

The third type of resonance does not occur in three or four particle clusters. Fig. 7(a) shows this resonance type for a seven-particle cluster, again occurring between two modes having mixed polarization but unlike the second type, creating a bi-directional instability. This is clearly indicated in the power spectra by an increase in energy density at mode numbers 9 and 10 and 17 and 18. Inspecting the $E_z'$-fr functions (Fig. 7(b)), deviations from the decoupled mode function shapes can also be clearly seen with the two different modes reconnected at lower and higher frequencies, leading to a gap in $E_z'$, corresponding to the instability region.

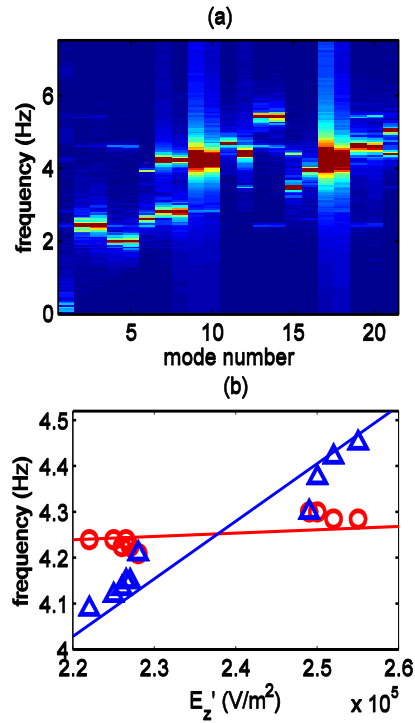

Fig. 7. (Color online) The third type of resonance as illustrated by (a) the power spectra and (b) $E_z'$-fr functions around the resonance point for one of the resonances seen in a seven particle cluster. The blue triangles and red circles represent the coupled vertical and horizontal modes, respectively. The decoupled shapes of the $E_z'$-fr functions are represented by the solid lines.

The characteristics shown above for the $E_z'$-fr functions around the resonance point for the second and third types of instability are similar to the positive and negative mode couplings seen in large crystals [17]. However, this should only be considered an analogy since in the present case, finite clusters and mode frequencies as a function of the external confinement are being investigated, i. e. the experimental environment is being allowed to change. Mode couplings in large crystals are usually investigated employing their dispersion relations, i.e. $k$-$\omega$ functions for a given experimental environment.

The $E_z'$-fr functions for three- to seven-particle clusters are shown in Figs. 8 and 9 for all modes and the range of variation in $E_z'$ employed in the simulations. As expected, horizontal mode frequencies remain constant while vertical mode frequencies increase as $E_z'$ increases. It is interesting to note that a lower limit for $E_z'$ must be established in order to maintain the 2D nature of the system; if $E_z'$ is below this limit, clusters tilt or split into two or more layers. In Figs. 8 and 9, the intersection points between coupled modes corresponding to the first, second and third type of resonances as defined above are marked by squares, triangles and circles, respectively. As discussed above, the first and third type of resonances both induce instabilities, as seen in Fig. 8(d), (e), (f) and Fig. 9(c), (d). In this case, the average particle velocities in the horizontal and vertical directions are plotted as functions of $E_z'$. Each peak corresponds to an instability occurring at the same value of $E_z'$ as a resonance point (intersection of the mode function) of either the third or first type.

For three and four particle clusters, only the first and second types of resonances occur. For the simplest case of three particles, the results shown are in agreement with previous research [33], where a generalized interparticle potential was assumed rather than a specific model for the ion wakefield. For particle numbers $\geq 5$, instabilities induced by the third type of resonance occur. Note that for particle numbers $\geq 5$, resonances between some coupled modes, e. g. the horizontal sloshing and the corresponding vertical mode, are not observed because the cluster loses its 2D nature before the two $E_z'$-fr function curves cross one another.

It can be seen in Fig. 9(c) and (d) that the instability peaks induced by the first type of resonance are much smaller than those induced by the third type, indicating that the coupling strength for the first type is weaker than that for the third type. This is consistent with the fact that no deviation from the decoupled $E_z'$-fr functions was detected for the first type of resonance. This can also cause the first type of instability to be indistinguishable within a particle velocity plot (Fig. 8(f)) when it occurs for a value of $E_z'$ which is close to that of a resonance of the third type.

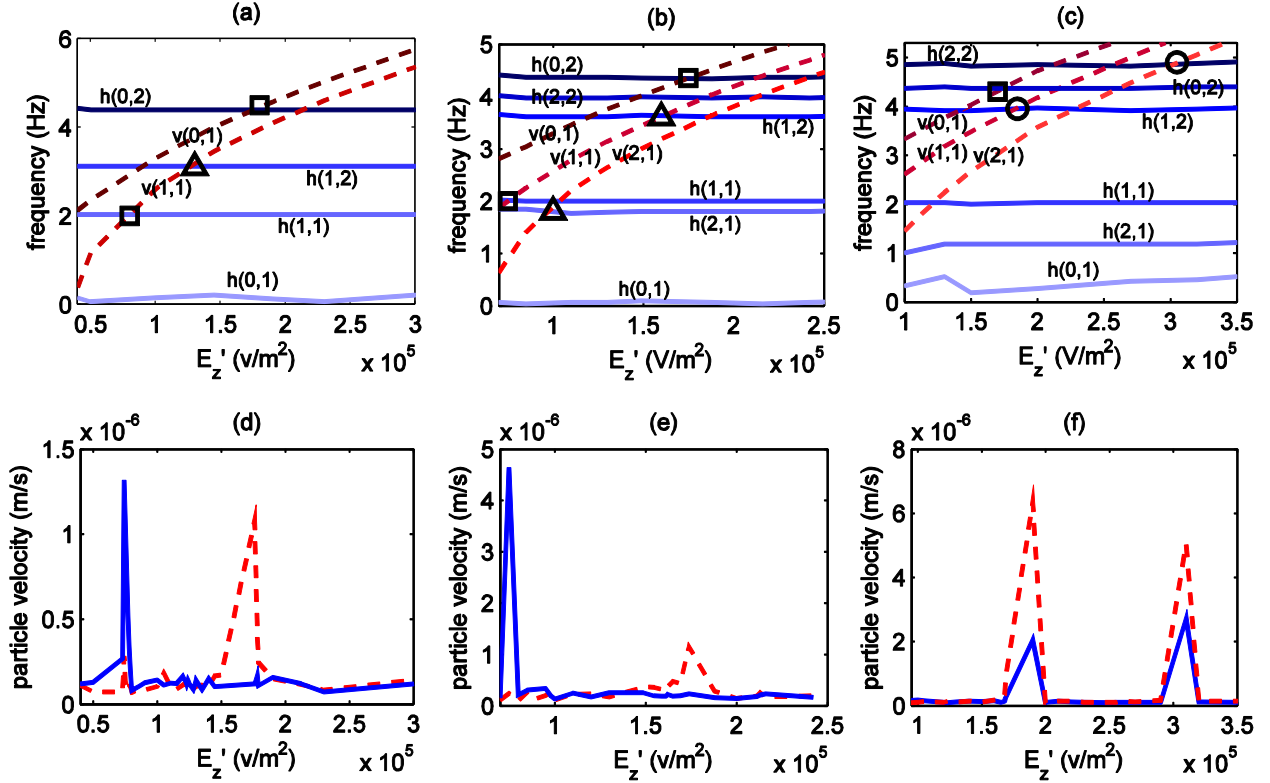

Fig. 8. (Color online) $E_z'$-fr functions for (a) 3, (b) 4, (c) 5 particle clusters and the average particle

velocity as a function of $E_z'$ for (d) 3, (e) 4, and (f) 5 particle clusters. In (a), (b), and (c), horizontal and vertical modes are represented by blue solid and red dashed lines, respectively. Intersection points corresponding to the first, second and third types of resonance are marked by squares, triangles and circles, respectively. In (d), (e) and (f), velocities in the horizontal and vertical directions are represented by blue solid and red dashed lines, respectively.

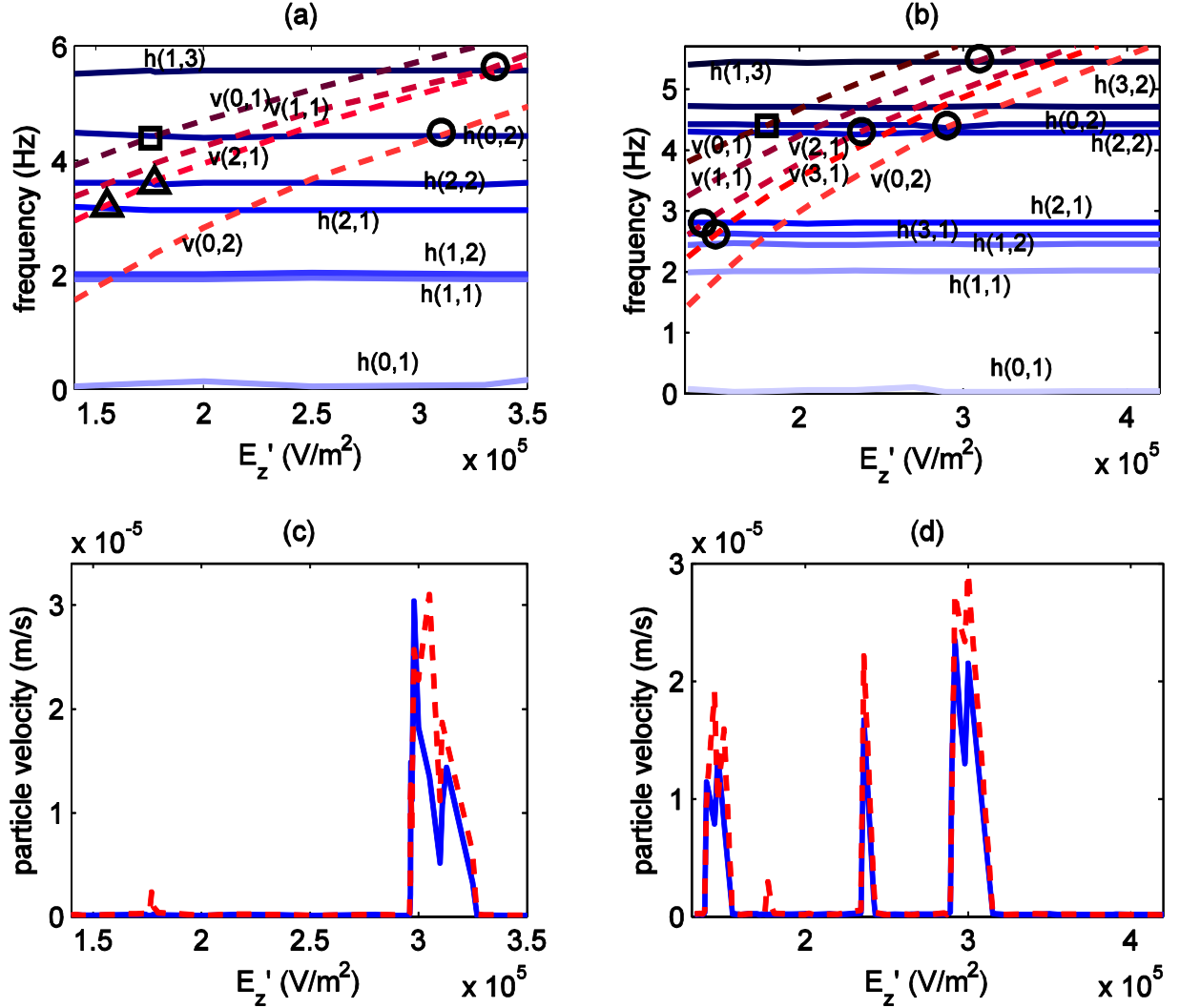

Fig. 9. (Color online) $E_z'$-$fr$ functions for (a) 6 and (b) 7 particle clusters and the average particle velocity as a function of $E_z'$ for (c) 6 and (d) 7 particle clusters. In (a-b), horizontal and vertical modes are represented by blue solid and red dashed lines, respectively. The intersection points corresponding to the first, second and third types of resonance are marked by squares, triangles and circles, respectively. In (c) and (d), velocities in the vertical and horizontal directions are represented by blue solid and red dashed lines, respectively.

## IV DISCUSSION

To determine the basic rules for mode coupling, first recall that vertical modes can be categorized using Bessel and trigonometric functions $J_m(k_{mn}r)e^{im\theta}$ with order integers $(m, n)$ in the circumferential and radial directions, as illustrated by Qiao and Hyde [28]. Horizontal dust cluster modes, on the other hand, have to date only been classified as shearlike or compressionlike modes by measuring the divergence and rotor of the particle velocity field [25, 26]. Nevertheless, there has been extensive research on the in-plane vibrations of continuous thin circular disks [55-59], which has shown that such eigenmodes can be represented by Bessel and trigonometric functions,

$$\phi(r,\theta,t) = A_m J_m(kr)\cos m\theta \sin\omega t$$
$$\psi(r,\theta,t) = B_m J_m(hr)\sin m\theta \sin\omega t \tag{3}$$

where $m$ is the order integer in the circumferential direction and $\phi$ and $\psi$ are defined by

$$u_r = \frac{1}{b}\left(\frac{\partial \phi}{\partial r} + \frac{1}{r}\frac{\partial \psi}{\partial \theta}\right)$$
$$u_\theta = \frac{1}{b}\left(\frac{1}{r}\frac{\partial \phi}{\partial \theta} - \frac{\partial \psi}{\partial r}\right) \tag{4}$$

where $b$ is the disk radius, and $u_r$ and $u_\theta$ are the radial and circumferential displacements, respectively. The constants $k$ and $h$ can be determined from the boundary conditions established for the disk, providing the other order integer $n$ in the radial direction and related to the vibration frequency $\omega$ by

$$k^2 = \frac{1}{2}\mu^{-1}(1-\nu)\rho\omega^2$$
$$h^2 = \mu^{-1}\rho\omega^2 \tag{5}$$

where $\mu$, $\upsilon$ and $\rho$ are Lame's constant, Poisson's ratio and mass density of the disk material, respectively.

Although the exact shape and frequency of the horizontal modes for finite dust clusters differ from those found for continuous disks, the modes can still be classified in the same manner using order integers ($m$, $n$). The order integer $m$ = 0, 1, 2, ... can be determined from the circumferential period, where for each $m$, the parameter $n$ = 1, 2, ... acts as the index of its frequency in ascending order since both $k$ and $h$ are proportional to the frequency $\omega$ based on the relationship in Eq. 5. Vertical modes can then be classified following the manner described in [28], where the order integers ($m$, $n$) are used as indices for the Bessel and trigonometric functions (eigenvectors for the normal modes), consistent with the classification method for horizontal modes. As a representative example, Fig. 10 illustrates the mode eigenvectors for a seven particle cluster, where each of the horizontal $h(m, n)$ and vertical $v(m, n)$ eigenvectors are marked and solid lines connect coupled modes.

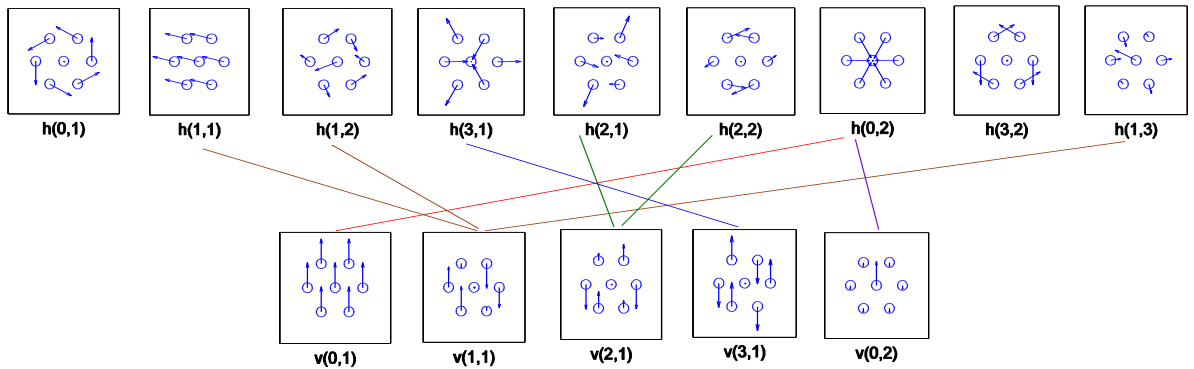

Fig. 10. (Color online) Mode eigenvectors for a seven particle cluster, where the horizontal (vertical) modes are shown at top (bottom) and solid lines connect coupled modes.

Inspecting Fig. 10, the following rules of coupling can be deduced. (1) Coupling only occurs between two modes having the same $m$. (2) Horizontal modes with dominating rotational characteristics, such as $h(0, 1)$ and $h(3, 2)$, are not coupled to a vertical mode. This implies that the coupling strength between two modes having the same $m$ but different $n$ is related to the horizontal mode's compressional and shear properties. To attain a better understanding of this, $h(2, 1)$ and $h(2, 2)$, both having the same $m$ and being coupled with $v(2, 1)$, were examined. Their divergence $\Psi_d$, rotor ($z$ component) $\Psi_r$ and the ratio $\Psi_d/\Psi_r$ were calculated following [25]. For $h(2, 1)$, the ratio $\Psi_d/\Psi_r \approx 0.41$ and for $h(2, 2)$, $\Psi_d/\Psi_r \approx 0.71$; thus, $h(2, 1)$ has a higher contribution from shear motion than does $h(2, 2)$. The strength of the coupling of $h(2, 1)$ and $h(2, 2)$ with $v(2, 1)$ can be manifested by the width of the corresponding instability peaks (Fig 9(d)). These instability peaks are shown in greater detail in Fig 11, where the maximum of the two peaks has been aligned. As can be clearly seen, the peak width for the $h(2, 1)v(2, 1)$ resonance is less than that found for the $h(2, 2)v(2, 1)$ resonance; hence, the postulate that higher shear contribution leads to weaker coupling. (The peak width is considered here instead of the height due to the large error involved in determination of the maximum velocity created by particle velocity fluctuation. The values of the vertical confinements $E_z'$ bounding the instability can be determined much more accurately). This postulate is also consistent with rule number two since modes which are not coupled to vertical mode, $h(0, 1)$ and $h(3, 2)$, are characterized by very low ratios $\Psi_d/\Psi_r$ of 0.003 and 0.04 respectively. This postulate may also be related to the fact that in infinite crystals, the in-plane shear wave does not couple to the out-of-plane wave [17, 21]. However, it must again be noted that since the number of resonance points examined is limited by the number of particles considered, this should only be considered a postulate until verified by further research.

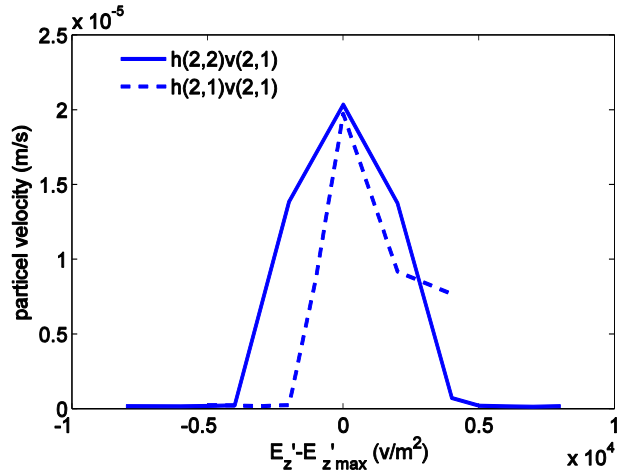

Fig. 11. (Color online) The instability induced by the resonances $h(2, 1)v(2, 1)$ and $h(2, 2)v(2, 1)$ in a seven particle cluster, indicated by peaks in the vertical velocities. The maximum of the two peaks are aligned to show the difference in peak width.

Whether a specified resonance will induce instability depends not only on the mode type but also on the specific structure and particle number of the cluster. For example, the resonance between $h(2, 2)$ and $v(2, 1)$ induces an instability in a seven particle cluster, while the resonance between the same two modes does not cause instability in a six particle cluster. This conclusion is further supported by calculations using the dynamic matrix of the cluster system (for details of this method, see [34, 35]), which shows that the dynamic matrix for a seven particle cluster has two complex conjugate eigenvalues corresponding to the modes $h(2, 2)$ and $v(2, 1)$, while all eigenvalues remain real for a six particle cluster.

Finally, it has been shown experimentally that the resonance between horizontal longitudinal and vertical transverse waves can cause nonequilibrium melting in large plasma crystals [20, 21]. This research shows that such resonance instabilities between coupled modes can cause dust cluster melting, as manifested by a solid-liquid transition, i. e. the transition from a state where all particles vibrate around their equilibrium positions to a state where particles are "hopping" between their equilibrium positions. . This microscopic view has been employed in many cases in research on finite system melting [23, 24, 30, 31, 37, 41-43] and its agreement with the definition in terms of the (thermodynamic) phase space [41] can be seen in [37]. The first type of resonance does not cause melting, but instead causes the system to vibrate as a whole. All instabilities induced by the third type of resonance can cause melting provided the coupling strength is large enough. The coupling strength can be "tuned" numerically to examine this in greater detail by varying the point charge $q$. For example, for a seven particle cluster, the system exhibits enhanced vibration at $q = Q/8$, but will not melt for values of $q < Q/4$.

A particularly simple and transparent quantity that is often used to characterize the solid and liquid states of a finite particle system is the Lindemann ratio $u_L$ defined as the magnitude of the particle fluctuation around equilibrium positions normalized to the interparticle distance [37]. For a 2D system, due to its logarithmic divergence with system size [60], a modified Lindemann ratio, defined by the relative interparticle distance fluctuations (IDF) is usually used [61, 62],

$$u_{rel} = \frac{2}{N(N-1)} \sum_{1 \leq i \leq j}^{N} \sqrt{\frac{\langle r_{ij}^2 \rangle}{\langle r_{ij} \rangle^2} - 1} \tag{6}$$

where $r_{ij}$ is the distance between particles $i$ and $j$, N is the particle number and $\langle ... \rangle$ denotes thermal averaging.

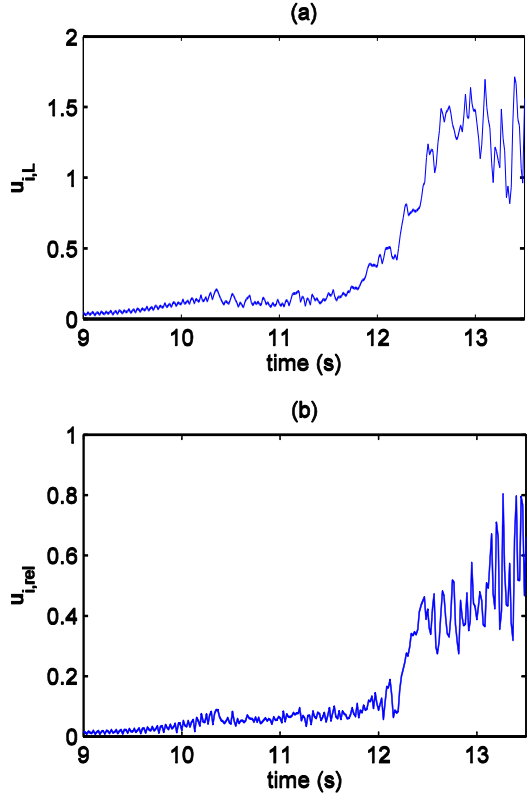

Fig. 12. (Color online) (a) $u_{i,L}$ and (b) $u_{i,\text{rel}}$ as a function of the frame number (time) for the melting of a seven particle cluster induced by the $h(2,1)v(2,1)$ resonance instability.

In order to investigate the temporal evolution of cluster melting, the instantaneous $u_{i,L}$ and $u_{i,\text{rel}}$ are defined by

$$u_{i,L} = \frac{1}{N}\sum_{i=1}^{N}\frac{\left|\vec{r_i}-\vec{r_{i0}}\right|}{a} \quad (7)$$

where $\vec{r_i}$ and $\vec{r_{i0}}$ are the instantaneous and equilibrium positions of particle $i$ and $a$ is the equilibrium interparticle spacing,

$$u_{i,rel} = \frac{2}{N(N-1)}\sum_{1\leq i\leq j}^{N}\frac{\left|r_{ij}-r_{ij0}\right|}{r_{ij0}} \quad (8)$$

where $r_{ij0}$ is the equilibrium distance between particles $i$ and $j$. Fig. 12 shows $u_{i,L}$ and $u_{i,\text{rel}}$ as functions of time for the melting of a seven particle cluster induced by the $h(2,1)v(2,1)$ resonance instability. Both $u_{i,L}$ and $u_{i,\text{rel}}$ show that melting proceeds via a two-step process (See Supplemental Material Movie1). In the first step, particles exhibit regular oscillations with their velocity vectors oriented in the direction of the corresponding mode's eigenvectors and oscillation amplitude growing with time. In the second step, cluster vibrations become irregular and eventually exhibit hopping of particles between their equilibrium positions, which is indicated by the sudden increase in $u_{i,L}$ and $u_{i,\text{rel}}$ at around 12.2 s. This two-step process is similar to that seen for the melting of a 2D dust cluster induced by a single particle underneath the crystal as reported by Ivanov and Melzer [43]. At a critical threshold value, $u_{i,\text{rel}}$ shows a sharp increase which always occurs at the time when a particle in the cluster begins to hop between equilibrium points, thus marking the solid-liquid transition. The corresponding increase in $u_{i,L}$ is more gradual, making the threshold value hard to distinguish (Fig. 12(a)). Table I lists threshold values for both $u_{i,L}$ and $u_{i,rel}$ at the solid-liquid transition for the nine instabilities induced by the third type of resonance for five to seven particle clusters. These values were obtained by averaging the values of $u_{i,L}$ and $u_{i,rel}$ over the ten frames before particle hopping first occurred. Table I shows that threshold values for $u_{i,rel}$ exhibits much less fluctuation (and exhibit smaller error bars) than $u_{i,L}$, in agreement with the temporal functions, indicating $u_{i,rel}$ to be a better marker than $u_{i,L}$ for characterization of the solid-liquid transition. The lowest threshold of $u_{i,rel}$ obtained is 0.1, corresponding to intermittent hopping between equilibrium particle positions (See Supplemental Material Movie2), while thresholds of 0.11 to 0.14 result from continuous hopping and a quick destruction of the stable cluster structure (See Supplemental Material Movie1). These values are in excellent agreement with the Lindemann criterion for macroscopic Coulomb systems (e.g. [61]). Therefore, nonequilibrium melting of dust clusters driven by resonance instabilities, although exhibiting a totally different mechanism from equilibrium melting driven by temperature increase [42], also closely obeys the Lindemann criterion.

Table I. Threshold values of $u_{i,L}$ and $u_{i,rel}$ at the solid-liquid transition for instabilities induced by the third type of resonance.

| *number of particles* | *mode types* | $u_L(\pm 0.05)$ | $u_{rel}(\pm 0.02)$ |
|---|---|---|---|
| 7 | $h(2,1)v(2,1)$ | 0.17 | 0.12 |
| 7 | $h(3,1)v(3,1)$ | 0.24 | 0.13 |
| 7 | $h(2,2)v(2,1)$ | 0.17 | 0.13 |
| 7 | $h(0,2)v(0,2)$ | 0.20 | 0.13 |
| 7 | $h(1,3)v(1,2)$ | 0.14 | 0.12 |
| 6 | $h(0,2)v(0,2)$ | 0.18 | 0.13 |
| 6 | $h(1,3)v(1,1)$ | 0.17 | 0.10 |
| 5 | $h(1,2)v(1,1)$ | 0.25 | 0.14 |
| 5 | $h(2,2)v(2,1)$ | 0.17 | 0.11 |

## V. CONCLUSIONS

The normal modes for three- to seven-particle 2D dust clusters formed within a complex plasma were obtained through numerical simulation. The ion wakefield was taken into account, employing the common point charge model, and representing a continuous energy input from the ambient plasma.

When compared to dust clusters acting only under a simple Debye–Hückel interaction potential (without an ion wakefield), the introduction of the ion wakefield is found to induce coupling between the horizontal and vertical modes. Resonance occurs when two coupled modes have the same frequency and can be classified into one of three types. The first type is a unidirectional resonance with one of the coupled modes being a sloshing mode. It creates a unidirectional instability in which only the energy density of the sloshing mode increases. The second type of resonance occurs between two modes having mixed polarization with a reconnection of the $E_z'$-$fr$ functions in the $E_z'$ regime and produces a forbidden frequency gap. It does not cause any instability. The third type of resonance also acts between two modes having mixed polarization, but includes a reconnection of the $E_z'$-$fr$ functions in the frequency regime leading to a gap in $E_z'$, corresponding to a bi-directional instability in which the energy density of both modes is increased. Due to the discrete nature of the normal mode frequencies, the system exhibit discrete instabilities as the vertical confinement is varied. Instabilities caused by the first type of resonance are found to be weaker than those caused by the third type, suggesting weaker coupling strengths.

Rules governing the coupling described above were determined by classifying mode eigenvectors employing Bessel and trigonometric functions indexed by order integers ($m$, $n$). It was shown that (1) Coupling only occurs between two modes having the same $m$, and that (2) horizontal modes with dominating rotational characteristics are not coupled to a vertical mode. Rule number two can be extended to provide the postulate that horizontal modes having higher shear contribution exhibit weaker coupling with vertical modes.

Finally, it was shown that all instabilities induced by the third type of resonance can cause melting, provided the coupling strength is large enough, while the first type of resonance only causes the system to vibrate as a whole. Melting was observed to proceed through a two-step process. In the first step, particles exhibit regular oscillations resulting in vibration in the direction of the corresponding mode eigenvectors, and oscillation amplitudes that increase with time. In the second step, vibrations become irregular causing particles to "hop" between equilibrium points, indicating a solid-liquid transition. The instantaneous relative interparticle distance fluctuations (IDF) $u_{i,\text{rel}}$ were found to act as a good marker for characterizing the solid-liquid transition in that its temporal evolution shows a sharp increase occurring when the first particle begins to hop between equilibrium points. The threshold value of $u_{i,\text{rel}}$ of 0.1 is in agreement with the Lindemann criterion for macroscopic Coulomb systems.

The results given here for simulated dust clusters resemble recent experimental results obtained for larger dust crystals [20-22]. Mode couplings, resonance instabilities and melting have all been clearly observed and the power spectra obtained for conditions near instabilities resembles the experimentally observed high energy density signature at the intersection of the horizontal and vertical dispersion relationships. At the same time, characteristics unique to small, finite systems have also been found, e.g. the discrete resonance instabilities observed for finite dust clusters. This prediction of discreteness should be relatively easy to examine experimentally, considering that typical experimental values were chosen for particle charge, mass, screening length, and horizontal confinement. Varying vertical confinement values can be easily achieved by changing the plasma power [44]. This will be discussed in an upcoming publication.

## ACKNOWLEDGMENTS

This research was supported in part by the NSF Grant No. 1002637.